# Order $\lambda^3$ Parameterization of Neutrino ( e, $\mu$, $\tau$, $\tau$ ' ) "Flavor" Oscillations in a Simplified SM4 Model and Associated $V_{CKM}$, $V_{MNS}$, and $V_{BIMAX}$ Matrixes

By

Henry Makowitz, Ph.D.
11800 South Country Club Drive
Idaho Falls, Idaho 83404-7848 USA
Email: Hmakowitz@GoBigWest.com
Tel:   ( 208 ) 524-2719
Fax:  ( 206 ) 350-2678

## ABSTRACT

Arguments are presented based on particle phenomenology and the requirement for Unitarity for a complex valued postulated four generation CKM Matrix ( $V_{CKM}$ ) based on a Sequential Fourth Generation Model (sometimes named SM4). A modified four generation QCD Standard Model Lagrangian is utilized per SM4. A four generation neutrino mass mixing MNS Matrix, ( $V_{MNS}$ ) is estimated utilizing a Unitary ( to $\theta$ ( $\lambda^k$ ) , k = 1, 2, 3, 4, … ) 4 x 4 Bimaximal Matrix, $V_{BIMAX}$. The Unitary $V_{BIMAX}$ is based on a weighted 3 x 3 $V_{BIMAX}$ scheme and is studied in conjunction with the postulated four generation $V_{CKM}$ complex Unitary Matrix. A single parameter $\lambda$ has been utilized in our analysis along with three complex $\Delta_{ij}$ phases. A four generation Wolfenstein Parameterization of $V_{CKM}$ is deduced which is valid for order $\lambda^3$. Experimental implications of the model are discussed. The issues of Baryogenesis in the context of Leptogenesis associated with MNS Matrix neutrino mixing and Baryogenesis associated with CKM Matrix quark mixing are discussed. These issues are studied in the context of the postulated complex four generation CKM Matrix and resulting complex MNS Matrix, predicted CP violating parameters, and a Fourth Generation Neutrino mass bound of $\geq$ 50 GeV. Our analysis is valid to and includes order $\lambda^3$ terms. Our work is a mathematical analysis of the consequences of such model.

## DISCUSSION

We will utilize the Sequential Fourth Generation Model ( sometimes named SM4 ) [ 1 , 2 ] in the discussion below without attempting much physical justification, only referring the reader to the cited references above as a source for justification. Presently, no physical data exists to validate any Fourth Generation models, hence our discussion is purely theoretical. We assume that neutrinos are Dirac particles.

As in [ 1, 3 ] we assume a Four Generation model given by Table I, below. Here the mass data is obtained from Ref. [ 4 ]. The quark entries in Table I consist of quark and anti-quark pair sets. Along with the assumptions discussed in "Note ( a )" this is a SM4 Model. We began our initial analysis with this model in Ref. [ 5 ]. In this paper we generalize our initial analysis and study a complete Unitary formulation. Our work is a mathematical analysis of the consequences of such a model.

### Table I -   Four Generation Model

| | $q = + 2/3$ | $q = - 1/3$ | charged leptons | neutral leptons |
|---|---|---|---|---|
| $1^{ST}_{GEN} \rightarrow$ | u ( 0.004 GeV/c$^2$ ) | d ( 0.008 GeV/c$^2$ ) | $e \pm$ | $\bar{\nu}(e), \nu(e)$ |
| $2^{ND}_{GEN} \rightarrow$ | c ( 1.35 GeV/c$^2$ ) | s ( 0.13 GeV/c$^2$ ) | $\mu \pm$ | $\bar{\nu}(\mu), \nu(\mu)$ |
| $3^{RD}_{GEN} \rightarrow$ | t ( 176 GeV/c$^2$ ) | b ( 4.7 GeV/c$^2$ ) | $\tau \pm$ | $\bar{\nu}(\tau), \nu(\tau)$ |
| $4^{TH}_{GEN} \rightarrow$ | t' ( >176 GeV/c$^2$ ) | b' ( >4.7<<t' GeV/c$^2$ ) | $\tau' \pm$ | $\bar{\nu}(\tau'), \nu(\tau')$ |

We have discussed the current experimental state of affairs elsewhere [ 5 ] and restate that there is no experimental motivation for a fourth generation model at this time and that our discussion is purely theoretical. One can argue that the fourth generation events are rare or that they occur at higher energies than what has been experimentally studied to date. Per references [ 1 , 2, 3 and  5 ] we will keep an open mind and pursue a simplified mathematical analysis of a SM4 model.

## ANALYSIS

Starting with order of magnitude arguments presented in [ 3 ] and the requirements for unitarity of a postulated four generation CKM Matrix ( $V_{CKM}$ ) [ refer to the 2004 review by F. J. Gilman et. al., in Ref. [ 4 a ] for the three generation CKM Matrix ] and utilizing a modified four generation QCD Standard Model Lagrangian$^{(a)}$ we obtain an expression for  $V_{CKM}$ given by Eq. ( 1.0 a), below. We express $V_{CKM}$ in the lowest order expansion in $\lambda$, the Wolfenstein parameter [ 4 a, 6 ]. Here $\lambda = \sin \vartheta_c \cong 0.2243 \pm 0.0016$ and $\vartheta_c \equiv$ Cabibbo mixing angle [ 4 a ]. One can view Eq. ( 1.0 a) as a hypothesis or postulate since no fourth generation angle data is available for comparison or motivation as was in [ 6 ].

**Eq. ( 1.0 a) below is a generalization of the $V_{CKM}$ that we used in Ref. [ 5 ].**

$$\begin{bmatrix} 1 & \lambda & \lambda^3 & \lambda^4 \\ -\lambda & 1 & \lambda^2 & \lambda^3 \\ \lambda^3 & -\lambda^2 & 1 & b_2\lambda \\ \lambda^4 & -\lambda^3 & b_1\lambda & 1 \end{bmatrix} = V_{CKM} \qquad \text{Eq. (1.0 a)}$$

**If $V_{CKM}$ is Unitary than by definition; $(V_{CKM})^+ = (V_{CKM})^{-1}$ where**

$(V_{CKM})^+ = (V_{CKM})^{T*}$ **here : T implies the Transpose ; * the complex conjugate**

**and $(V_{CKM})^+ \times (V_{CKM}) = (V_{CKM}) \times (V_{CKM})^+ = 1$**

**$V_{CKM}$ for the three generation system [ 4, 6 ] is given in the Wolfenstein Parameterization by Eq. (1.0 b), below;**

**Eq. (1.0 b),**

$$\begin{bmatrix} 1-\lambda^2/2 & \lambda & A\lambda^3(\rho - i\eta) \\ -\lambda & 1-\lambda^2/2 & A\lambda^2 \\ A\lambda^3(1-\rho - i\eta) & -A\lambda^2 & 1 \end{bmatrix} = V_{CKM} + \vartheta(\lambda^4)$$

**where $i = \sqrt{-1}$.**

**We will study the four generation $V_{CKM}$ in more detail below for $\vartheta(\lambda^3)$. We will generalize our previous results.**

**$V_{CKM}$ for the four generation system is given in general by Eq. (1.0 c), below, if we assume the form of Eq. (1.0 a) and Eq. (1.0 b) above ;**

Eq. (1.0 c),

$$\begin{vmatrix} 1 - \lambda^2/2 & \lambda & A\lambda^3(\rho - i\eta) & C_0\lambda^4 \\ -\lambda & 1 - \lambda^2/2 & A\lambda^2 & C_3\lambda^3 \\ A\lambda^3(1 - \rho - i\eta) & -A\lambda^2 & 1 \times C_5 & (a_3 + a_4 i)\lambda \\ C_1\lambda^4 & -C_2\lambda^3 & (a_1 + a_2 i)\lambda & (1 + C_4) \end{vmatrix} = V_{CKM} + \vartheta(\lambda^4)$$

Here the $C_i$'s are real or complex, the $a_i$'s are real, and A, $\rho$, and $\eta$ are real and have the usual Wolfenstein coefficient values.

Unitary matrix requirements constrain Eq. (1.0 c) for $\vartheta(\lambda^3)$ by Eq. (1.0 d), below;

$$(V_{CKM}) \times (V_{CKM})^\dagger = (V_{CKM})^\dagger \times (V_{CKM}) = \delta_{i,j} + \vartheta(\lambda^4) \qquad \text{Eq. (1.0 d)}$$

These constraints result with[b] $C_5 = (1 - \lambda^2)$, $C_4 = (-\lambda^2)$, $C_3 = Ai$, $C_2 = -A$, $a_4 = a_2 = 1.0$, $a_3 = 1.0$, and $a_1 = -1.0$. Since we are only interested in $\vartheta(\lambda^3)$ results, $C_1 = C_0 \equiv 1.00$ are arbitrary in this work. The reader should note that our formulation has resulted in three complex phases for the four generation $V_{CKM}$.

Let us consider an order of magnitude analysis first. We will find later on that this is sufficient. For Eq. (1.0 a) we obtain $V_{CKM})^{-1}$, given in Appendix I with standard matrix algebra. Here for $b_1 = b_2 = \sqrt{-1} \equiv i$; $b_3 = b_4 = -i$ a Unitary solution results. (Note that this system also allow the real valued Unitary solution: $b_1 = -1$, $b_2 = 1$; $b_3 = 1$, $b_4 = -1$ and complex valued forms ($\pm 1 + i$) with their complex conjugates per Eq. (1.0 c));

$$\begin{vmatrix} 1 & -\lambda & 0 & 0 \\ \lambda & 1 & -\lambda^2 & 0 \\ 0 & \lambda^2 & 1 & b_4 \lambda \\ 0 & 0 & b_3 \lambda & 1 \end{vmatrix} = (V_{CKM})^{-1} \qquad \text{Eq. (2.0)}$$

In this paper we are interested in the; $b_1 = b_2 = \sqrt{-1} \equiv i$ ; $b_3 = b_4 = -i$ simple Unitary complex valued solution and not the real valued system studied previously [ 5 ]. Using the complex valued forms ( $\pm 1 + i$ )$\lambda$ will not change our results.

Simple relations between the Neutrino and Quark Mixing Angles are discussed in Ref. [ 7 ]. A summary of the "Global Analysis" of Neutrino Data can be found in Ref. [ 8 ]. More complex relationships between Neutrino and Quark Mixing Angles can be found in Refs. [ 9, 10, 11, 12 ]. We will use the $V_{BIMAX}$ which is discussed in Refs. [ 9, 10, 11 ], extensively analyzed in Ref. [ 10 ] and utilized in Ref. [ 5 ]. Although other relationships have also been proposed ( re: Ref. [ 12 ] ) we will utilize $V_{BIMAX}$ from Ref. [ 10 ] as our initial analysis starting point. The neutrino mass mixing MNS Matrix, ( $V_{MNS}$ ) [ refer to the 2005 review by B. Kayser in Ref. [ 4 a ] for the three generation MNS Matrix ] will be estimated utilizing the Bimaximal Matrix discussed in the literature [ 10 ], $V_{BIMAX}$ and our postulated four generation $V_{CKM}$.

$$V_{MNS-A} = (V_{CKM})^{-1} \times V_{BIMAX} \qquad \text{Eq. (3.0 a)}$$

or

$$V_{MNS-B} = V_{BIMAX} \times (V_{CKM})^{-1} \qquad \text{Eq. (3.0 b)}$$

The reader should note that equations (3.0 a) and (3.0 b ) do not commute.

We have generalized $V_{BIMAX}$ ( 3 x 3 ) to a ( 4 x 4 ) matrix as per our discussion in Appendix II and [ 5 ]. Note that our analysis in [ 5 ] was not Unitary but is Unitary in this work to $\theta$ ( $\lambda^k$ ) , k = 1, 2, 3, 4, … . We utilize this Unitary ( to low order ) 4 x 4 Bimaximal Matrix, $V_{BIMAX}$ based on a weighted 3 x 3 $V_{BIMAX}$ scheme discussed in Appendix II, per Eq. (B 4.0). We will state the results in general form. We will only discus $V_{MNS-A}$ given by Eq. (3.0 a) in this paper :

$$\begin{vmatrix} B_1 \times (\sqrt{2}/2) & B_1 \times (\sqrt{2}/2) & 0 & (+B_2) \\ B_1 \times (-1/2) & B_1 \times (1/2) & B_1 \times (\sqrt{2}/2) & (-B_2) \\ B_1 \times (1/2) & B_1 \times (-1/2) & B_1 \times (\sqrt{2}/2) & (+B_2) \\ (+B_2) & (-B_2) & (+B_2) & (-B_5) \end{vmatrix} = V_{BIMAX} \qquad \text{Eq. (4.0)}$$

where;

$B_2 = +[1 - B_1^2]^{1/2}$

( Pg. - 5 - )

$B_5 = +[1 - 3(1 - B_1^2)]^{1/2}$

$B_1 = +[(n/m)]^{1/2}$ where $n < m$, and $n, m$ are non zero positive integers
and $(n/m) < 1$, $(n/m) > 2/3$

$$V_{BIMAX} + \lambda \times \begin{vmatrix} +B_1/2 & -B_1/2 & -B_1\sqrt{2}/2 & +B_2 \\ +B_1\sqrt{2}/2 & +B_1\sqrt{2}/2 & 0 & +B_2 \\ -iB_2 & +iB_2 & -iB_2 & +iB_5 \\ -iB_1/2 & +iB_1/2 & -iB_1\sqrt{2}/2 & -iB_2 \end{vmatrix} + \theta(\lambda^2) = V_{MNS-A} \quad \text{Eq. (5.0)}$$

The reader should refer to Appendix III for the real valued $V_{MNS-A}$ expression.
Note again that our analysis in [ 5 ] was not Unitary and is Unitary in this work.

For order unitarity of $V_{BIMAX}$ to $\theta(\lambda)$ ; $B_1 = 0.9487$, $B_2 = 0.3162$, $B_5 = 0.8367$.
For order unitarity of $V_{BIMAX}$ to $\theta(\lambda^2)$ ; $B_1 = 0.9950$, $B_2 = 0.1000$, $B_5 = 0.9849$.
For order unitarity of $V_{BIMAX}$ to $\theta(\lambda^3)$ ; $B_1 = 0.9995$, $B_2 = 0.0316$, $B_5 = 0.9985$.
For order unitarity of $V_{BIMAX}$ to $\theta(\lambda^4)$ ; $B_1 = 0.999998$, $B_2 = 0.0020$, $B_5 = 0.999994$.

What we mean by this is that order unitarity of $V_{BIMAX}$ to $\theta(\lambda^4)$ implies;

$(V_{BIMAX}) \times (V_{BIMAX})^+ = (V_{BIMAX})^+ \times (V_{BIMAX}) = \delta_{i,j} + \vartheta(\lambda^4)$

and this implies that $B_2 = 0.00$ for $\vartheta(\lambda^3)$.

Since $(V_{CKM})^{-1} \times (V_{CKM}) = (V_{CKM}) \times (V_{CKM})^{-1}$ to $\theta(\lambda^2)$ and the product is Unitary for $\theta(\lambda)$ per Appendix I ( i.e., $(1 + \lambda^2)$ diagonal elements, zero off diagonal elements to $\theta(\lambda^3)$ ), $V_{BIMAX}$ to $\theta(\lambda^2)$ appears to be sufficient for our current approximation of $V_{MNS-A}$. The reader needs to refer the "<u>MASS AND ENERGY BOUNDS</u>" section of this paper for further analysis and considerations of this issue and EW Bounds on our analysis. EW Bounds also significantly effect the required precision of the current analysis.

Neutrino Oscillations are described as a "Quantum Mechanical Process" and are discussed in [ 2005 review by B. Kayser in Ref. [ 4 a ] ]. Summarizing the results of [ 4 a ] , if $\alpha = e, \mu, \tau, \tau$ ', are the lepton "flavors" for the neutrino eigenstate $|\nu_\alpha>$, and $|\nu_i>$ the neutrino "mass" eigenstates for $i = 1, 2, 3, 4,\ldots$ etc. where we have assumed $i \leq 4$, L is the distance from the initial neutrino starting location, $m_i$ is the neutrino mass and $E_i$ is the neutrino energy for a generalization to four generations we obtain from Ref. [ 4 a ] ;

Eq. (6.0 a):

$$|\nu_\alpha(L)\rangle \approx \sum_i (V_{\alpha,i})^* \exp(-i(m_i^2/2E_i)L) |\nu_i\rangle$$

and

Eq. (6.0 b)

$$|\nu_\alpha(L)\rangle \approx \sum_\beta \left[\sum_i (V_{\alpha,i})^* \exp(-i(m_i^2/2E_i)L) V_{\beta,i}\right] |\nu_\beta\rangle$$

**Note that the $V_{i,j}$ Matrix elements in Eq. ( 6.0 ) are the MNS lepton mass mixing matrix components [2005 review by B. Kayser in Ref. [ 4 a ] ]. We see from Eq. ( 6.0 ) that an initial single mass state and flavor neutrino will become a superposition of all MNS Matrix neutrino mixed mass and flavor states at a distance L.**

**The complex terms ( i.e., $b_1 = b_2 = \sqrt{-1} = i$ ) in our model $V_{CKM}$ would imply, if the usual theoretical interpretation is utilized for $V_{CKM}$, $\theta(\lambda)$ CP violating Matrix elements. This is much larger than the usual $\theta(\leq \lambda^3)$ CP violating terms found in the Wolfenstein Parameterization [ 4 a, 6 ] of $V_{CKM}$ for the 3 x 3 system. Such large CP terms, if they exist may have a major impact on our understanding of Baryogenesis. CP and C violations are required to explain the matter-antimatter asymmetry [ 13 ] associated with the Baryogenesis Problem along with Baryon Number Violating Processes [ 13 ] and that the processes that produce the baryon asymmetry should take place out of thermal equilibrium assuming symmetrical initial conditions [ 13 ]. This issue needs extensive experimental investigation. For recent reviews of the Baryogenesis Problem the reader should refer to Refs. [ 14 , 15, 16 ]. The reader should refer to this paper's section entitled "<u>CP VIOLATION</u>" for a further analysis and discussion.**

**The $V_{MNS-A}$ neutrino mixing Matrix given by Eq. ( 5.0 ) above contains $\theta(\lambda)$ and $\theta(\lambda^2)$ complex elements along with real components. If one interprets the complex elements as CP violating coefficients, again much lower order in $\lambda$ results are obtained than if one used only the $\theta(\leq \lambda^3)$ CP violating terms found in the Wolfenstein Parameterization [ 4 a, 6 ] of the 3 x 3 $V_{CKM}$ Matrix and utilized real coefficients for the fourth generation of the 4 x 4 $V_{CKM}$ Matrix. In our low order formulation the CP Violating terms occur only for the third and fourth rows of the MNS Matrix. However our low order analysis is only valid to $\theta(\lambda^2)$. The reader needs to refer the "<u>MASS AND ENERGY BOUNDS</u>" section of this paper for further analysis and considerations of this issue and EW Bounds on our analysis. EW Bounds also significantly effect the required precision of the current analysis. The reader should also refer to this paper's section entitled "<u>CP VIOLATION</u>" for a further analysis and discussion as well as Appendix IV and Appendix V.**

**CP violating neutrino oscillations can lead to a solution of the Baryogenesis Problem [ 15 and 16 ] usually referred to as Leptogenesis. We do not attempt to perform an extensive review of this area of research here. However we refer the reader to a number of publications on this subject that are relevant [ 17, 18, 19, 20, 21, 22, 23 ]. To our knowledge no work has been done on our MNS Matrix SM4 based system of equations or similar SM4 systems resulting from BIMAX and CKM Marices in the context of Leptogenesis. We plan to investigate Leptogenesis further with our theory. We are aware of some preliminary four generation lepton analysis that differs from our results [ 23 ].**

## MASS AND ENERGY BOUNDS

**Based on b' data ( lack of observation ) in Ref. [ 4 a ] and [ 4 b ] one concludes that; Mass Limits for b' ( 4 th Generation ) Quark or Hadron in p+p- Collisions are given by;**

      **>190 GeV   at 95 % CL  ( ACOSTA )  not seen**
      **>199 GeV   at 95 % CL  ( AFFOLDER ) not seen**
      **>128 GeV   at 95 % CL  ( ABACHI ) not seen**

**Mass Limit for b' ( 4 th Generation ) Quark or Hadron in e+e- Collisions is given by;**

      **>46 GeV    at 95 % CL  ( DECAMP ) not seen**
      **96 – 103 GeV   Ref. [ 24 ]  not seen**

**We suggest that the rest mass M ( fourth generation quark b' ) < Mass ( W Boson). b' mass > than the bottom quark but << than the top and t' quark. The data range above for b' candidates suggests for p+p- Collisions this as a possibility at < 5 % CL. For e+e- Collisions the possibility for the existence of b' is >> 5% CL at a mass of >46 GeV except for between 96 – 103 GeV per Ref. [ 24 ].**

**Heavy Fourth Generation Neutrino mass bounds have the most data scatter in Ref. [ 4 a ] and [ 4 b ] ranging in values from >10, >39.5, >45, to >2400 GeV. ( not observed ) Also, Z-lineshape analysis [ 25 ] excludes the 4 th generation at a 95 % CL for a neutrino mass < 47 GeV. Theoretical arguments suggest the > 5 GeV bound as a starting point for more careful experimental analysis per Refs. [ 26 and 27 ] ( "Lee-Weinberg" stable heavy neutrino cosmological lower bound ). However, a Fourth Generation Neutrino mass bound of ≥ 50 GeV is more consistent with Z-lineshape constraints per Refs. [ 25 and 28 ] and LEP [ 25 ] experimental analysis.**

**A region of fourth generation parameter space in agreement with all experimental constraints and minimal contributions to the electroweak ( EW ) precision oblique data is discussed and analyzed in Ref. [ 29 ]. Applying this analysis to our work constrains $B_2$ to a value of ≤ 0.02. Our $V_{CKM}$ is consistent with the requirements of the analysis in Ref. [ 29 ]. However our analysis of $V_{MNS}$ requires further discussion.**

For order unitarity of $V_{BIMAX}$ to $\theta(\lambda^3)$; $B_1 = 0.9998$, $B_2 = 0.02$, $B_5 = 0.9994$. However $V_{BIMAX}$ unitarity requirements to $\theta(\lambda^4)$; $B_1 = 0.999998$, $B_2 = 0.0020$, $B_5 = 0.999994$ imply that $B_2 = 0.00$ for $\vartheta(\lambda^3)$. Hence, the $V_{BIMAX}$ matrix unitarity requirements are more constraining than EW requirement in our formulation of $V_{BIMAX}$.

So far in this paper we have presented low order ( to $\theta(\lambda^2)$ ) results. Hence, so far in this work we need to take $B_2 = 0.00$ and $B_1$ as well as $B_5$ of $\theta(1.00)$. This approximation results in a non-trivial $V_{MNS-A}$ formulation since it is valid for $B_2 = 0.00$ for $\vartheta(\lambda^3)$ per the discussion above.

In general, we need to extend our analysis up to $\theta(\lambda^3)$ for $B_1$ and $B_5$. We perform this analysis in Appendix IV and Appendix V. Since we have obtained a four generation $V_{CKM}$ for $\vartheta(\lambda^3)$ per Eq. (1.0 c) this is just simple matrix algebra with $B_2 = 0.00$ for $V_{BIMAX}$ and results in $V_{MNS-A}$ for $\theta(\lambda^3)$. We find that EW requirements are satisfied in this higher order analysis.

## CP VIOLATION

In this paper a single Wolfenstein parameter $\lambda$ has been utilized in our analysis as apposed to the usual three complex $\Delta_{ij}$ phases and six real angles [ 30 ], greatly simplifying the work[c]. The appearance of three complex $\Delta_{ij}$ phases for a four generation $V_{CKM}$ model introduces additional terms of $\theta(\leq \lambda^3)$ that are taken into consideration in Appendix IV and Appendix V. This analysis of higher order terms in $V_{MNS-A}$ includes an expansion in $B_1$ and $B_5$. It should be noted that the three generation $V_{CKM}$ case only results in one complex phase [ 4 b, 6, and 30 ].

CP Violations are reviewed in texts [ 31, 32, 33 ]. For the three generation $V_{CKM}$ case a single invariant, J, the Jarlskog invariant [ 34, 35, 36, 37 ] results as an area of the triangles formed by $(V_{CKM})^+ \times (V_{CKM}) = (V_{CKM}) \times (V_{CKM})^+ = 0.00$. This area in the Wolfenstein parameterization is given below [ per Ref. 4 b ] as J / 2 ;

$$J \cong \lambda^6 A^2 \eta \qquad \text{Eq. (7.0)}$$

If $J = 0.00$ there is no CP Violation. If $J \neq 0.00$ CP Violation exists, however the analysis of its magnitude is system specific. For a greater number of generations than three other invariants arise that are similar to J. The reader should refer to the analysis presented in Refs. [ 36, 37, 38, 39, 40, 41 and 42 ] for four generations and beyond.

For three generations, text [ 32 ] makes the general observation that triangles with equal sides of $\theta(\lambda^3)$ and area of $\theta(\lambda^6)$ corresponding to B decay lead to large CP violation vs. K decay which results in "spear-like" thin triangles of area $\theta(\lambda^6)$ that result in small CP violations. If we assume that this "general observation" is correct we can make the following statements about our four generation $V_{CKM}$ system;

Approximate triangles with sides of $\theta(\lambda^3)$ exists for fourth generation mixing. These systems are quadrangles resulting in approximate triangles and should result in large CP Violation. They are given below;

$$V_{cd}(V_{t'd})^* + V_{cs}(V_{t's})^* + V_{cb}(V_{t'b})^* + V_{cb'}(V_{t'b'})^* = 0.00 \qquad \text{Eq. (8.0 a)}$$

$$\theta(\lambda^5) \quad \theta(\lambda^3) \quad \theta(\lambda^3) \quad \theta(\lambda^3)$$

$$V_{t'd}(V_{cd})^* + V_{t's}(V_{cs})^* + V_{t'b}(V_{cb})^* + V_{t'b'}(V_{cb'})^* = 0.00 \qquad \text{Eq. (8.0 b)}$$

$$\theta(\lambda^5) \quad \theta(\lambda^3) \quad \theta(\lambda^3) \quad \theta(\lambda^3)$$

In addition two other families of quadrangles may be of interest since their maximal angle areas are also $\theta(\lambda^6)$. These quadrangles also mix in the fourth generation and their sides are of order;

$$\theta(\lambda^7) + \theta(\lambda^5) + \theta(\lambda) + \theta(\lambda) = 0.00 \qquad \text{Eq. (9.0 a)}$$

and

$$\theta(\lambda^4) + \theta(\lambda^2) + \theta(\lambda^2) + \theta(\lambda^4) = 0.00 \qquad \text{Eq. (9.0 b)}$$

We leave the study of CP Violations of the $V_{MNS}$ system for future work.

## Summary and Conclusions

Arguments have been presented based on particle phenomenology and the requirements for Unitarity of a postulated four generation CKM Matrix ($V_{CKM}$) in a simplified hypothetical SM4 Model. A ( 4 x 4 ) complex $V_{CKM}$ Matrix has been formulated utilizing order of magnitude arguments suggested in [ 3 ] and Matrix Unitarity constraints. A modified four generation QCD Standard Model Lagrangian has been utilized [a]. A single Wolfenstein parameter $\lambda$ has been utilized in our analysis as apposed to the usual three complex $\Delta_{ij}$ phases and six real angles [ 30 ], greatly simplifying the work[c]. An Order $\lambda^3$ parameterization has been analyzed. In this analysis the four generation $V_{CKM}$ results in the usual number of three complex phases per the standard literature [ 30 ].

We evaluated $V_{MNS-A} = (V_{CKM})^{-1} \times V_{BIMAX}$ utilizing a Unitary 4 x 4 Bimaximal Matrix, $V_{BIMAX}$ based on a weighted 3 x 3 $V_{BIMAX}$. A formal mathematical structure has been formulated to study the effect of the weighting on four generation neutrino mixing and the weighting has been evaluated for $V_{BIMAX}$ as a function of Unitary ( for $\theta(\lambda^k)$, k = 1, 2, 3, 4 ). The limiting behavior of the $V_{BIMAX}$ Matrix weighting was investigated and a set of weighting parameters were selected so that $\theta(\lambda)$ and $\theta(\lambda^3)$ accuracy could be achieved for the analysis based on $V_{BIMAX}$ Matrix Unitarity weighting requirements. We performed this study utilizing a complex valued Unitary 4 x 4 $V_{CKM}$ model Matrix.

Our specific hypothetical[c] complex Unitary 4 x 4 $V_{CKM}$ model Matrix results in $\theta(\lambda)$ CP violating matrix elements. This is much larger than the usual $\theta(\leq \lambda^3)$ CP violating terms found in the Wolfenstein Parameterization [ 4 a, 6 ] of $V_{CKM}$ for the 3 x 3 system. Such large CP violating terms, if they exist could have a major impact on our understanding of Baryogenesis as well as the Leptogenesis scenario for Baryogenesis. CP Violating terms of $\theta(\lambda)$ are predicted for both CKM Matrix quark mixing and MNS Matrix neutrino mixing and are a consequence of Unitary requirements for the CKM Matrix. These issues needs extensive experimental and further theoretical investigation. Our work should be viewed as a mathematical and theoretical analysis of the consequences of a possible, but not unique hypothetical Unitary $\theta(\lambda^3)$ SM4 model.

**Notes: ( a ) In SM4, due to symmetry reasons we postulate two 4 th generation
quarks, b' ( q= -1/3 ) and t' ( q= 2/3 ), and t' mass > t mass.
Due to our desire to remain within the quark model ( and QCD
formalism of the Standard Model ) we utilize the standard
QCD Lagrangian given by Eq. 9.1, 9.2 and 9.3 contained in [ 2005
review, Quantum Chromodynamics and its Coupling, I. Hinchliffe,
Ref. [ 4 a ] ], with eight quark specific flavors as apposed to the usual six.
Each quark flavor comes in three colors, and gluons come in eight color
combinations. Hadrons are color singlet combinations of quarks, anti-
quarks and gluons. Now the CKM Matrix will be a 4 x 4 Matrix as
apposed to the usual 3 x 3 CKM Matrix.**

**( b ) In the Wolfenstein three generation parameterization we only modify
the CKM{ t, b } matrix element by a factor of ( $1 - \lambda^2$ ). This is justified
within the experimental knowledge of the CKM{ t, b } element ( per the
lbl pdg tables and discussed experimental data within, Ref. [ 4 ] ).**

**( c ) We have assumed that the six real angles produce a negligibly small total
effect on our results, and can be represented by one real angle, $\lambda$ .
Although this may appear to be unphysical an not theoretically
justifiable ( since there is no available data on the fourth generation ),
due to Note ( b ) above, the effect should be of $\leq \vartheta(( 1 - \lambda^2 ))$ as a
best guess estimate. However, the reader should note that our
CKM Matrix parameterization is not unique ( refer for example to
Ref. [ 42 ] ). Our CKM Matrix parameterization should be treated as
a mathematically allowed solution to the physical problem due to it's
unitarity for $\vartheta ( \lambda^3 )$ .**

## Appendix I

**For the $V_{CKM}$ Matrix given by Eq. ( A 1.0 ) below:**

$$\begin{bmatrix} 1 & \lambda & \lambda^3 & \lambda^4 \\ -\lambda & 1 & \lambda^2 & \lambda^3 \\ \lambda^3 & -\lambda^2 & 1 & b_2\lambda \\ \lambda^4 & -\lambda^3 & b_1\lambda & 1 \end{bmatrix} = V_{CKM} \quad \text{Eq. ( A 1.0 )}$$

**we obtain $(V_{CKM})^{-1}$, given by Eq. ( A 2.0 ) below with standard matrix algebra.**

$$\begin{bmatrix} 1 & -\lambda & 0 & 0 \\ \lambda & 1 & -\lambda^2 & 0 \\ 0 & \lambda^2 & 1 & b_4\lambda \\ 0 & 0 & b_3\lambda & 1 \end{bmatrix} = (V_{CKM})^{-1} \quad \text{Eq. ( A 2.0 )}$$

**For $b_1 = b_2 = \sqrt{-1} \equiv i$ ; $b_3 = b_4 = -i$** 	Eq. ( A 3.0 )

$(V_{CKM})^+ = (V_{CKM})^{T*} = (V_{CKM})^{-1}$ to $\theta(\lambda^2)$. ( Low order Unitary )

**Note also that for Eqn. ( A 3.0 );**

$$\begin{bmatrix} (1+\lambda^2) & 0 & 0 & 0 \\ 0 & (1+\lambda^2) & 0 & 0 \\ 0 & 0 & (1+\lambda^2) & 0 \\ 0 & 0 & 0 & (1+\lambda^2) \end{bmatrix} + \theta(\lambda^3) = (V_{CKM}) \times (V_{CKM})^{-1} \quad \text{Eq. ( A 4.1 )}$$

**and** $(V_{CKM})^{-1} \times (V_{CKM}) = (V_{CKM}) \times (V_{CKM})^{-1}$ to $\theta(\lambda^2)$	Eq. ( A 4.2 )

**For $b_1 = -1$, $b_2 = 1$ ; $b_3 = 1$, $b_4 = -1$**  Eq. ( A 5.0 )

$(V_{CKM})^+ = (V_{CKM})^{T*} = (V_{CKM})^{-1}$ to $\theta(\lambda^2)$. ( Low order Unitary )

**For completeness we present the results for $V_{MNS-A}$ associated with this case in Appendix III.**

**For $b_1 = b_2 = 1$ ; $b_3 = b_4 = -1$**  Eq. ( A 6.0 )

$$\begin{bmatrix} (1+\lambda^2) & 0 & 0 & 0 \\ 0 & (1+\lambda^2) & 0 & 0 \\ 0 & 0 & (1-\lambda^2) & 0 \\ 0 & 0 & 0 & (1-\lambda^2) \end{bmatrix} + \theta(\lambda^4) = (V_{CKM}) \times (V_{CKM})^{-1} \quad \text{Eq. ( A 7.0 )}$$

and $(V_{CKM})^{-1} \times (V_{CKM}) = (V_{CKM}) \times (V_{CKM})^{-1}$ to $\theta(\lambda^2)$  Eq. ( A 8.0 )

Note that $(V_{CKM})^+ = (V_{CKM})^{T*} \neq (V_{CKM})^{-1}$ for this case ( Not Unitary ).

( Pg. - 17 - )

## Appendix II

We have been able to formulate a 4 x 4 $V_{BIMAX}$ that is Unitary to lowest order and that has the same properties as the original 3 x 3 case [ 5 ] ;

$V_{BIMAX} \Rightarrow \sum_{row} |V(i,j)|^2 = 1$ is true as well as $V_{BIMAX} \Rightarrow \sum_{COLUMN} |V(i,j)|^2 = 1$.

This $V_{BIMAX}$ is given by Eq. ( B 1.0 ), below:

$$\begin{bmatrix} (\sqrt{3}/2) \times (\sqrt{2}/2) & (\sqrt{3}/2) \times (\sqrt{2}/2) & 0 & (1/2) \\ (\sqrt{3}/2) \times (-1/2) & (\sqrt{3}/2) \times (1/2) & (\sqrt{3}/2) \times (\sqrt{2}/2) & (-1/2) \\ (\sqrt{3}/2) \times (1/2) & (\sqrt{3}/2) \times (-1/2) & (\sqrt{3}/2) \times (\sqrt{2}/2) & (1/2) \\ (1/2) & (-1/2) & (1/2) & (-1/2) \end{bmatrix} = V_{BIMAX} \quad \text{Eq. (B 1.0)}$$

The reader should also note that for Eq. ( B 1.0 );

$(V_{BIMAX})^+ \times (V_{BIMAX}) \neq (V_{BIMAX}) \times (V_{BIMAX})^+$ \quad\quad Eq. (B 2.0)

This result can be generalized to a "weighted" $V_{BIMAX}$ Matrix which utilizes the initial 3 x 3 $V_{BIMAX}$ [ 5 ] as the starting point as follows:

$$\begin{bmatrix} B_1 \times (\sqrt{2}/2) & B_1 \times (\sqrt{2}/2) & 0 & (B_2) \\ B_1 \times (-1/2) & B_1 \times (1/2) & B_1 \times (\sqrt{2}/2) & (B_3) \\ B_1 \times (1/2) & B_1 \times (-1/2) & B_1 \times (\sqrt{2}/2) & (B_4) \\ (B_6) & (B_7) & (B_8) & (B_5) \end{bmatrix} = V_{BIMAX} \quad \text{Eq. (B 3.0)}$$

The conditions $V_{BIMAX} \Rightarrow \sum_{row} |V(i,j)|^2 = 1$ as well as $V_{BIMAX} \Rightarrow \sum_{COLUMN} |V(i,j)|^2 = 1$ result with eight equations and assuming the same sign conventions as adapted in Eq. (B 1.0) we obtain:

$$\begin{vmatrix} B_1 \times (\sqrt{2}/2) & B_1 \times (\sqrt{2}/2) & 0 & (+B_2) \\ B_1 \times (-1/2) & B_1 \times (1/2) & B_1 \times (\sqrt{2}/2) & (-B_2) \\ B_1 \times (1/2) & B_1 \times (-1/2) & B_1 \times (\sqrt{2}/2) & (+B_2) \\ (+B_2) & (-B_2) & (+B_2) & (-B_5) \end{vmatrix} = V_{BIMAX} \quad \text{Eq. (B 4.0)}$$

where;

$B_2 = +[1 - B_1^2]^{1/2}$

$B_5 = +[1 - 3(1 - B_1^2)]^{1/2}$

$B_1 = +[(n/m)]^{1/2}$ where $n < m$, and n, m are non zero positive integers

and $(n/m) < 1$, $(n/m) > 2/3$

Utilizing Eq. (B 4.0) we have obtained generalized expressions for;

$(V_{BIMAX})^+ \times (V_{BIMAX}) = U_1(i,j)(4 \times 4)$  Eq. (B 5.0)

and

$(V_{BIMAX}) \times (V_{BIMAX})^+ = U_2(i,j)(4 \times 4)$  Eq. (B 6.0)

where

$U_1(i,j)(4 \times 4) \neq U_2(i,j)(4 \times 4)$  Eq. (B 7.0)

Numerical results indicate low order unitarity to $\theta(\lambda)$ for $V_{BIMAX}$ given by Eq. (B 4.0) with an optimum value of (n/m) and the resulting coefficients $B_1$, $B_2$, and $B_5$. This order unitarity to $\theta(\lambda)$ exists for $(n/m) \approx (9/10)$. Order unitarity to $\theta(\lambda^2)$ exists for $(n/m) \approx (99/100)$ and order unitarity to $\theta(\lambda^3)$ exists for $(n/m) \approx (999/1000)$.

## Appendix III

**For $b_1 = -1$, $b_2 = 1$; $b_3 = 1$, $b_4 = -1$**            Eq. ( C 1.0 )

$(V_{CKM})^+ = (V_{CKM})^{T*} = (V_{CKM})^{-1}$ to $\theta(\lambda^2)$. ( Low order Unitary )

**This results for $V_{MNS-A}$ with Eq. ( C 2.0 ) below:**

$$V_{BIMAX} + \lambda \times \begin{vmatrix} +B_1/2 & -B_1/2 & -B_1\sqrt{2}/2 & +B_2 \\ +B_1\sqrt{2}/2 & +B_1\sqrt{2}/2 & 0 & +B_2 \\ -B_2 & +B_2 & -B_2 & +B_5 \\ +B_1/2 & -B_1/2 & +B_1\sqrt{2}/2 & +B_2 \end{vmatrix} + \theta(\lambda^2) = V_{MNS-A}$$

                                                                                             Eq. ( C 2.0 )

## Appendix IV

**Starting with Eq. (1.0 a) in the main text of this work and for:**

**$b_1 = -1$, $b_2 = 1$ ;**                                                                                              Eq. ( D 1.0 )

$$V_{MNS-A} = (V_{CKM})^+ \times V_{BIMAX}$$        Eq. ( D 2.0 )

**For and $B_2 = 0.00$, per EW and Unitarity constraints:**

$$\begin{vmatrix} B_1 \times (\sqrt{2}/2) & B_1 \times (\sqrt{2}/2) & 0 & 0 \\ B_1 \times (-1/2) & B_1 \times (1/2) & B_1 \times (\sqrt{2}/2) & 0 \\ B_1 \times (1/2) & B_1 \times (-1/2) & B_1 \times (\sqrt{2}/2) & 0 \\ 0 & 0 & 0 & (-B_5) \end{vmatrix} = V_{BIMAX} \quad \text{Eq. ( D 3.0 )}$$

**This results for $V_{MNS-A}$ with Eq. ( D 4.0 ) below:**

$$V_{BIMAX} + \lambda \times \begin{vmatrix} +B_1/2 & -B_1/2 & -B_1\sqrt{2}/2 & 0 \\ +B_1\sqrt{2}/2 & +B_1\sqrt{2}/2 & 0 & 0 \\ 0 & 0 & 0 & +B_5 \\ +B_1/2 & -B_1/2 & +B_1\sqrt{2}/2 & 0 \end{vmatrix} + $$

$$\lambda^2 \times \begin{vmatrix} 0 & 0 & 0 & 0 \\ -B_1/2 & +B_1/2 & -B_1\sqrt{2}/2 & 0 \\ -B_1/2 & +B_1/2 & +B_1\sqrt{2}/2 & 0 \\ 0 & 0 & 0 & 0 \end{vmatrix} + \quad \text{Eq. ( D 4.0 )}$$

$$\lambda^3 x \begin{vmatrix} +B_1/2 & -B_1/2 & +B_1\sqrt{2}/2 & 0 \\ 0 & 0 & 0 & +B_5 \\ +B_1\sqrt{2}/2 & +B_1\sqrt{2}/2 & 0 & 0 \\ -B_1/2 & +B_1/2 & +B_1\sqrt{2}/2 & 0 \end{vmatrix} + \theta(\lambda^4) = V_{MNS-A}$$

**It should be noted by the reader that $\lambda^3$ x $B_5$ $\cong$ 0.0113 versus the EW constraint of $\leq$ 0.02 for the $V_{MNS-A}$ ( 2, 4 ) th matrix element. Hence, both EW and Unitarity constraints are satisfied in our discussion, in this Appendix IV. However our formulation here for ( $V_{CKM}$ ) x ( $V_{CKM}$ )$^+$ is valid to only $\theta(\lambda^2)$ ( i.e., $\lambda$ ). We need to evaluate out current results further utilizing Eq. (1.0 c) with the constraints of Eq. (1.0 d). $\theta(\lambda^3)$ results for $V_{MNS-A}$ are presented in Appendix V.**

## Appendix V

**Starting with Eq. (1.0 c) in the main text of this work and for constraints;**

$$(V_{CKM}) \times (V_{CKM})^+ = (V_{CKM})^+ \times (V_{CKM}) = \delta_{i,j} + \vartheta(\lambda^4) \quad \quad \text{Eq. (E 1.0)}$$

**results with**[c] $C_5 = (1 - \lambda^2)$, $C_4 = (-\lambda^2)$, $C_3 = Ai$, $C_2 = -A$, $a_4 = a_2 = 1.0$, $a_3 = 1.0$, and $a_1 = -1.0$, $C_1 = C_0 \equiv 1.00$. **Again, the reader should note that our formulation has resulted in three complex phases for the four generation $V_{CKM}$ per Eq. (E 2.0), below:**

Eq. (E 2.0),

$$\begin{bmatrix} 1 - \lambda^2/2 & \lambda & A\lambda^3(\rho - i\eta) & \lambda^4 \\ -\lambda & 1 - \lambda^2/2 & A\lambda^2 & Ai\lambda^3 \\ A\lambda^3(1 - \rho - i\eta) & -A\lambda^2 & 1 - \lambda^2 & (1+i)\lambda \\ \lambda^4 & A\lambda^3 & (-1+i)\lambda & 1 - \lambda^2 \end{bmatrix} = V_{CKM} + \vartheta(\lambda^4)$$

**Here A, $\rho$, and $\eta$ are real and have the usual Wolfenstein coefficient values. ( A = 0.80, $\rho$ = 0.19, and $\eta$ = 0.36, Ref. [ 43 ] )**

$$V_{MNS-A} = (V_{CKM})^+ \times V_{BIMAX} = (V_{CKM})^{-1} \times V_{BIMAX} + \theta(\lambda^4) \quad \quad \text{Eq. (E 3.0)}$$

**For and $B_2 = 0.00$, per EW and Unitarity constraints:**

$$\begin{bmatrix} B_1 \times (\sqrt{2}/2) & B_1 \times (\sqrt{2}/2) & 0 & 0 \\ B_1 \times (-1/2) & B_1 \times (1/2) & B_1 \times (\sqrt{2}/2) & 0 \\ B_1 \times (1/2) & B_1 \times (-1/2) & B_1 \times (\sqrt{2}/2) & 0 \\ 0 & 0 & 0 & (-B_5) \end{bmatrix} = V_{BIMAX} \quad \quad \text{Eq. (E 4.0)}$$

This results for $V_{MNS-A}$ with Eq. ( E 5.0 ) below:

Eq. ( E 5.0 ),

$$V_{BIMAX} + \lambda \times \begin{bmatrix} +B_1/2 & -B_1/2 & -B_1\sqrt{2}/2 & 0 \\ +B_1\sqrt{2}/2 & +B_1\sqrt{2}/2 & 0 & 0 \\ 0 & 0 & 0 & (1+i)B_5 \\ (1-i)B_1/2 & -(1-i)B_1/2 & (1-i)B_1\sqrt{2}/2 & 0 \end{bmatrix} +$$

$$\lambda^2 \times \begin{bmatrix} -B_1\sqrt{2}/4 & -B_1\sqrt{2}/4 & 0 & 0 \\ (B_1/4 - AB_1/2) & (-B_1/4 + AB_1/2) & (-B_1\sqrt{2}/4 - AB_1\sqrt{2}/2) & 0 \\ (-B_1/2 - AB_1/2) & (B_1/2 + AB_1/2) & (-B_1\sqrt{2}/2 + AB_1\sqrt{2}/2) & 0 \\ 0 & 0 & 0 & B_5 \end{bmatrix} +$$

$$\lambda^3 \times \begin{bmatrix} A(1-\rho+i\eta)B_1/2 & -A(1-\rho+i\eta)B_1/2 & A(1-\rho+i\eta)B_1\sqrt{2}/2 & 0 \\ 0 & 0 & 0 & -AB_5 \\ A(\rho+i\eta)B_1\sqrt{2}/2 & A(\rho+i\eta)B_1\sqrt{2}/2 & 0 & 0 \\ AiB_1/2 & -AiB_1/2 & -AiB_1\sqrt{2}/2 & 0 \end{bmatrix}$$

$+ \theta(\lambda^4) = V_{MNS-A}$

It should be noted by the reader that $|-\lambda^3 A B_5| \cong 0.009$ versus the EW constraint of $\leq 0.02$ for the $V_{MNS-A}(2, 4)$ th matrix element. Hence, both EW and Unitarity constraints are satisfied in our discussion, in this Appendix V. Also, our formulation here for $(V_{CKM}) \times (V_{CKM})^+$ is valid to $\theta(\lambda^4)$ ( i.e., $\lambda^3$ ).

It should also be noted that a $\lambda^4$ analysis is required to obtain a non zero value for the $V_{MNS-A}(1, 4)$ th matrix element.